\documentstyle[titlepage,12pt]{article}
\begin{document}

\renewcommand{\theequation}{\thesection.\arabic{equation}}

\newcommand{\be}{\begin{equation}}
\newcommand{\ee}{\end{equation}}
\newcommand{\bea}{\begin{eqnarray}}
\newcommand{\eea}{\end{eqnarray}}

\title{Strings in Homogeneous Background Spacetimes}
\author{Mariusz P. D\c{a}browski 
\footnote{E-mail: mpdabfz@uoo.univ.szczecin.pl}\\
         {\it Institute of Physics, University of Szczecin, Wielkopolska 15, 
          70-451 Szczecin, Poland}
{\ }\\
Arne L. Larsen \footnote{E-mail: all@fysik.ou.dk}\\
         {\it Institute of Physics, University of Odense, Campusvej 55,
           5230 Odense M, Denmark}}
\date{\today}

\maketitle
\begin{abstract}
The string equations of motion for some homogeneous (Kantowski-Sachs, 
Bianchi I and Bianchi IX) background spacetimes are given, and solved 
explicitly in some simple
cases. This is motivated by the recent developments in string cosmology, where
it has been shown that, under certain circumstances, such spacetimes appear
as string-vacua. 

Both tensile and null strings are considered. Generally, it
is much simpler to solve for the null strings since then we deal with the null
geodesic equations of General Relativity plus some additional constraints. 

We consider in detail an ansatz corresponding to circular strings, and we
discuss the possibility of using an elliptic-shape string ansatz in the case of
homogeneous (but anisotropic) backgrounds.

\end{abstract}
\newpage

\section{Introduction}
\setcounter{equation}{0}

There have been quite a lot of interest during the last few years in the
evolution of strings in fixed curved backgrounds; for reviews, see for 
instance \cite{san}. It is well-known that the
string equations of motion in curved spacetimes form a complicated 
system of second-order non-linear
coupled partial differential equations which, in general, is non-integrable. 
However, there are some special string configurations for which the
equations are exactly solvable
(see for instance \cite{san} and references therein). 
A lot of such explicit solutions have already been found, but they
were restricted to just the most symmetric spacetimes such like Minkowski, de
Sitter, anti-de Sitter, Schwarzschild and Robertson-Walker ones. 

We shall consider both tensile and null (tensionless) strings. Generally, it
is much simpler to solve for the null strings \cite{null}, since then we deal 
with the null
geodesic equations of General Relativity plus some additional constraints. 
It might seem then, that the dynamics of null strings is quite trivial.
However, this is not true. Although each individual point along the null string
follows a null geodesic, the null string as a whole may experience highly 
non-trivial dynamics \cite{nullsol}. 
The situation is qualitatively similar to that
of congruences in General Relativity, 
that is, "bundles of rays": each ray in the bundle is just following a
geodesic, but the propagation of the bundle as a whole can be highly
non-trivial due to tidal forces, as described by the Raychaudhuri equation.

In this paper we extend the discussion of the tensile and null 
string evolution to the
homogeneous, but anisotropic, spacetimes of Kantowski-Sachs, Bianchi type I 
and IX. In fact, Kantowski-Sachs solutions with negative and zero curvature are
just axisymmetric Bianchi type I and III Universes. This means that 
only positive
curvature Kantowski-Sachs models are different from Bianchi type Universes. 

Our main motivation is the recent development in string-cosmology. It has been 
shown that Kantowski-Sachs and Bianchi type spacetimes, under 
certain circumstances, appear as string-vacua, i.e. as solutions
to the $\beta$-function equations \cite{fradcall} to 
some (low) order in $\alpha'.$ Thus, in this paper,
we consider the dynamics of a string moving in a condensate of the
massless string modes.

The paper is organized as follows. In Sec. II, we give the tensile and null 
string equations
of motion and constraints in generic curved spacetimes. 
In Sec. III, we specialise to the case of
positive curvature Kantowski-Sachs (K-S) backgrounds, in 
particular, we consider the K-S Universe with cosmological term and the K-S
Universe with stiff-fluid matter.
In both cases we solve the equations of motion for tensile and null strings 
after making appropriate ans\"{a}tze, and we give the physical
interpretation of the solutions.  In Sec. IV we turn to the Bianchi Universes.
We first discuss some of the problems concerning finding explicit tensile
string solutions in spacetimes of Kasner type, and secondly, 
we give some explicit 
null string solutions. Finally in Sec. V, we consider circular strings 
in axisymmetric Bianchi IX Universes, and in Sec. VI we give our conclusions.

\section{String Equations of Motion and Constraints}
\setcounter{equation}{0}

Let us consider the tensile and the null string 
equations of motion in a compact formula
\be
\ddot{X}^{\mu} + \Gamma^{\mu}_{\nu\rho} \dot{X}^{\nu}\dot{X}^{\rho} 
= \lambda \left( {X}^{\prime\prime\mu} + \Gamma^{\mu}_{\nu\rho} 
X^{\prime\nu}X^{\prime\rho} \right)   ,
\ee
where dot means derivative with respect to the string coordinate $\tau$ 
and prime means derivative with respect to the string coordinate $\sigma$.
The constraints read as 
\begin{eqnarray}
g_{\mu\nu}\dot{X}^{\mu}\dot{X}^{\nu} & = & - \lambda 
g_{\mu\nu}X^{\prime\mu}X^{\prime\nu}   ,\\
g_{\mu\nu}\dot{X}^{\mu}X^{\prime\nu} & = & 0   .
\end{eqnarray}
For $\lambda = 1$ we have the tensile strings while $\lambda = 0$ applies for 
the null strings. Notice that in Ref. \cite{expansion}, expansion schemes were
considered, essentially using $\lambda$ ($ \lambda \sim 1/\alpha^{\prime}$,
where $\alpha^{\prime}$ is the 
inverse string tension) as a continuous expansion 
parameter; in this paper we simply use $\lambda$ as a discrete 
parameter discriminating between tensile and null strings. 

From the above we can see that for the null 
strings, we have the null geodesic equation of General Relativity 
supplemented by the constraint 
(II.3). For both null and tensile strings
the invariant string size is defined by (closed string) 
\begin{equation}
S(\tau)=\int_0^{2\pi}\;S(\tau,\sigma)\;d\sigma,
\end{equation}
where
\begin{equation}
 S(\tau,\sigma)= \sqrt{-g_{\mu \nu}X^{\prime \mu}X^{\prime\nu}}\; .
\end{equation}

\section{Strings in Kantowski-Sachs Background}
\setcounter{equation}{0}

In this section, we 
will consider the string equations of motion in homogeneous 
Kantowski-Sachs spacetimes. The validity of the Kantowski-Sachs spacetime as 
being consistent string vaccuum (solution to the $\beta$-function
equations to the lowest order in $\alpha^{\prime}$)
is discussed elsewhere \cite{kslow}. Here we just 
concentrate on the motion of a test string in Kantowski-Sachs backgrounds. 

The Kantowski-Sachs spacetime is given by the metric \cite{ks}
\begin{equation}
ds^2=dt^2-A^2(t)dr^2-B^2(t)d\Omega _k^2,
\end{equation}
where the "angular" metric is 
\begin{eqnarray}
d\Omega _k^2 &=&d\theta ^2+S^2(\theta )d\varphi ^2,  \nonumber \\
S(\theta ) &=&\left\{ 
\begin{array}{l}
\sin {\theta }\hspace{0.5cm}{\rm for}\hspace{0.3cm}k=+1, \\ 
\theta \hspace{0.5cm}{\rm for}\hspace{0.3cm}k=0, \nonumber \\ 
\sinh {\theta }\hspace{0.5cm}{\rm for}\hspace{0.3cm}k=-1,
\end{array}
\right. \ 
\end{eqnarray}
and $A, B$ are the expansion scale factors. Here, $r\in \;]-\infty,\infty[\;$, 
while the range of $t$ depends on the particular cosmology. For $k = +1$ the 
coordinates $\theta$ and
$\varphi$ describe, as usual, the angles on the 2-sphere. 
Only $k = +1$ models
fall outside the Bianchi classification, but usually one refers to all three
curvature models as Kantowski-Sachs Universes. In this paper, we mainly
consider $k = +1$ models.

As a first example of a string configuration, we apply the following string 
ansatz 
\be
X^0 = t(\tau), \hspace{0.5cm}X^1 = r(\tau), \hspace{0.5cm}X^2 = \theta(\tau),
 \hspace{0.5cm}X^3 = \varphi = \sigma,   
\ee
which describes a circular string winding around the 2-sphere. The functions
$(t(\tau),r(\tau),\theta(\tau)),$ which describe the dynamics of the string, 
are to be determined from the equations of motion.

For the metric (III.1), we start with the string equations of motion (II.1) 
and constraints (II.2)-(II.3), which now reduce to 
\begin{eqnarray}
\ddot{t} + AA_{,t}\dot{r}^2 + BB_{,t} \dot{\theta}^2 - 
\lambda BB_{,t}\sin^2{\theta} & = & 0   ,\\
\ddot{r} + 2 \frac{A_{,t}}{A} \dot{t}\dot{r} & = & 0   ,\\
\ddot{\theta} + 2 \frac{B_{,t}}{B} \dot{t}\dot{\theta} + 
\lambda \sin{\theta}\cos{\theta} & = & 0   ,\\
\dot{t}^2 - A^2 \dot{r}^2 - B^2 \dot{\theta}^2 - \lambda B^2 \sin^2{\theta}
& = & 0   .
\end{eqnarray}
The Eq. (III.4) easily integrates to give
\be
\dot{r} = \frac{dr}{d\tau} =  \frac{k}{A^2}   ,
\ee
with $k =$ const. The other equations, in general, cannot be integrated,  
thus we must either consider special Kantowski-Sachs spacetimes or make
further restrictions for the ansatz. Notice, however, that
for both cases $\lambda = 0, 1,$ the invariant string size is given 
by 
\be
S(\tau) = 2\pi|B(\tau) \sin{\theta(\tau)}|   .
\ee

\subsection{Tensile Strings}

For the tensile strings, $\lambda = 1,$ and Eq.(III.5) is fulfilled 
automatically under the assumption that $\theta =$ const. $ = \pi/2$ in 
(III.2). In the rest of Subsection III.A, 
we restrict ourselves to this case.
Then after inserting (III.7) into (III.6), we obtain 
\be
\dot{t}^2 = \frac{k^2}{A^2} + B^2   ,
\ee
or, explicitly in terms of the string time coordinate,  
\be
\tau(t) = \int^{t}{\frac{|A|\;dt}{\sqrt{k^2 + A^2B^2}}}   ,
\ee
while from (III.7) we get
\be
r(t)=k \int^{t}{\frac{dt}{|A|\sqrt{k^2 + A^2B^2}} }   .
\ee
Notice also that (III.3) is automatically fulfilled now.

\subsubsection{$\Lambda$-term solutions}

First we refer to one of the simplest solutions given for the scale factors, 
which is the $k = +1$ Kantowski-Sachs Universe with only the cosmological term 
\cite{gron}. These are:
\begin{eqnarray}
A(t) & = & H_{0}^{-1} \sinh{H_{0} t}   ,\\
B(t) & = & H_{0}^{-1} \cosh{H_{0} t}   , 
\end{eqnarray}
with $H_{0} =$ const., and we consider only the expanding phase ($t\geq 0$)
of the Universe.
After inserting (III.12)-(III.13) into (III.10) we have (choosing boundary
conditions such that $t(0)=0$) 
\be 
\tau(t) =  H_{0} \int_0^{t} { \frac{\sinh{(H_{0}t)}}{\sqrt{k^2H_{0}^4 + 
 \sinh^2{(H_{0}t)} \cosh^2{(H_{0}t)}}}\; dt}   ,
\ee
and from (III.11) we obtain 
\be 
r(t) = k H_{0}^3 \int_{t_0}^{t} 
 \frac{dt}{\sinh{(H_{0}t)} \sqrt{k^2H_{0}^4 + 
 \sinh^2{(H_{0}t)} \cosh^2{(H_{0}t)}}}    .
\ee
By inverting (III.14), giving $t(\tau),$ Eq.(III.15) then gives 
explicitly $r(\tau).$
The Eqs.(III.14)-(III.15) can be transformed to the form of the standard
elliptic integrals. For instance, by making the substitution 
$\cosh{(H_{0}t)} = \sqrt{z},$ Eq.(III.14) becomes
\be
\tau(z) = \frac{1}{2} \int_1^{z} 
{\frac{dz}{\sqrt{z(z - z_{1})(z - z_{2})}}}   ,
\ee
with 
\be
z_{1,2} = \frac{1}{2} \left( 1 \pm \sqrt{1 - 4k^2 H_{0}^4}\; \right)   ,
\ee
and can thus be evaluated explicitly eventually 
yielding $t(\tau).$ However, the detailed form will not be important here.
The invariant
string size in this case is simply
\be
S(\tau)=2\pi H_0^{-1}\cosh(H_0t(\tau)),
\ee
i.e., the string size follows the expansion of the Universe. This is
easily understood, since the string is simply 
winding around the equatorial plane 
of the 2-sphere.  

\subsubsection{Time-symmetric stiff-fluid solutions}

The solutions for the scale factors for the time-symmetric $k = +1$
(recollapsing) stiff-fluid Kantowski-Sachs model is given by \cite{dab}
\begin{eqnarray}
A(t) & = & b     ,\\
B(t) & = & \frac{\sqrt{M}}{b} \sqrt{ 1 - \frac{b^2}{M} 
(t - t_{0})^2}\;      ,
\end{eqnarray}
with $(b, M, t_{0})$  constants. The constant $M$ appears in the density of
stiff-fluid matter conservation law $\rho A^2 B^4 = M$. The Kantowski-Sachs
model described by scale factors (III.19)-(III.20) begins and ends at "barrel"
singularities $(A =$ const., $B = 0$) \cite{col}. For simplicity we will 
take $t_{0} = \sqrt{M}/b\;$ from now on, so that the range of $t$ is
$\;t\in [0,2\sqrt{M}/b].\;$ 
For the exact solution (III.19)-(III.20), Eqs. (III.7)
and (III.10) integrate to give  
\bea 
t(\tau) & = & \frac{\sqrt{M}}{b}+\frac{\sqrt{k^2 + M}}{b}
\;\sin \left(\tau - 
\mbox{arcsin}\sqrt{\frac{M}{k^2 + M}}\;\right)   ,\\
r(\tau) & = & \frac{k}{b^2}\tau+r_{0}  ,
\eea
where we took boundary conditions such that $t(0)=0.$
Having this, one can express $B$ in terms of $\tau$, i.e.,
\be
B(\tau) = \frac{\sqrt{M}}{b} \sqrt{1 - \frac{k^2 + M}{M} 
\sin^2 \left(\tau - 
\mbox{arcsin}\sqrt{\frac{M}{k^2+M}}\;\right)}  ,
\ee
which in the limit $k = 0$ ($r =$ const.) gives 
\be
B(\tau) = \frac{\sqrt{M}}{b}\sin{\tau} .
\ee
According to (III.8), the invariant string size is simply $S(\tau) =
2\pi B(\tau),$ which 
means that the string trivially starts with zero size, then
expands to a maximum size and finally contracts to zero size 
again together with the Universe, i.e. at 
$\tau=2\mbox{arcsin}\sqrt{M/(k^2+M)}.$ 
This again comes from the fact that the
string simply winds around the equatorial plane of the 2-sphere. 

In the next subsection, we shall consider more complicated (null) string 
solutions with non-trivial dynamics.

\subsection{Null strings}

The above solutions for tensile strings were all obtained for $\theta=\pi/2.$
However,  
for the null strings, $\lambda = 0,$ we can easily integrate (III.5) ,
still keeping the general form of $\theta = \theta(\tau)$ in the ansatz 
(III.2), to obtain 
\be
\dot{\theta} = \frac{d\theta}{d\tau} = \frac{l}{B^2}   ,
\ee
with $l =$ const.
The Eq.(III.6) now becomes 
\be
\dot{t}^2 = \frac{k^2}{A^2} + \frac{l^2}{B^2}   ,
\ee
or, explicitly
\be
\tau(t) =  \int^{t}{\frac{|AB|\;dt}{\sqrt{l^2A^2 + k^2B^2}} }   ,
\ee
while (III.7) becomes
\be
r(t) = k \int^{t}{\frac{|B|\;dt}{|A|\sqrt{l^2A^2 + k^2B^2}} }   .
\ee               
\subsubsection{$\Lambda$-term solutions}

Inserting (III.12)-(III.13) into (III.27), we have 
an exact relation between the spacetime and the string time coordinates 
\begin{equation}
H_{0} t(\tau) = {\rm arccosh}\sqrt{\frac{l^2+[H^{2}_{0}(k^2 + 
l^2) \tau + |k|]^2}{k^2 + l^2}}   ,
\end{equation}
where we choose again boundary conditions such that $t(0)=0.$
Having given (III.29), one can write down (III.12)-(III.13) in terms 
of the $\tau$-coordinate as 
\begin{eqnarray}
A(\tau) & = & \frac{1}{H_{0}} \sqrt{\frac{-k^2+[H_{0}^2(k^2 + 
l^2)\tau +|k|]^2}{k^2 + l^2}}   ,\\
B(\tau) & = & \frac{1}{H_{0}} \sqrt{\frac{l^2+[H_{0}^2(k^2 + 
l^2)\tau + |k|]^2}{k^2 + l^2}}   .
\end{eqnarray}
This allows us to integrate (III.7) and (III.25) to give
\begin{eqnarray}
r(\tau) & = & -\frac{k}{|k|}
{\rm arccth}\left(1+\frac{H_{0}^2}{|k|}(k^2 + l^2)\tau \right) + 
r_{0}   ,\\
\theta(\tau) & = &  -\frac{l}{|l|}
{\rm arctg}\left(\frac{|k|}{|l|}+\frac{H_{0}^2}{|l|}
(k^2 + l^2)\tau \right) + 
\theta_{0},
\end{eqnarray}
with $r_{0}, \theta_{0} =$ const. 
The invariant string size (III.8) is given by 
\bea
S(\tau)&=&2\pi|B(\tau)\sin\theta(\tau)|\\
&=&\frac{2\pi}{H_0\sqrt{k^2+l^2}}
\left| \Big( |k|+H_0^2(k^2+l^2)\tau\Big)\sin\theta_0
- l\cos\theta_0\right|. \nonumber
\eea
It is useful to consider some special cases. For $l = 0$ (i.e., 
$\theta = \theta_{0}$),  we have 
\begin{eqnarray}
A(\tau) & = &  \sqrt{H_{0}^2k^2\tau^2 +2|k|\tau}   ,\\
B(\tau) & = & H_{0} |k| \tau +H_0^{-1}  ,\\
t(\tau) & = & \frac{1}{H_{0}} {\rm arccosh}{(H_{0}^2 |k| \tau+1)}   ,\\
r(\tau) & = & -\frac{k}{|k|}{\rm arccth}{(H_{0}^2|k|\tau+1)} + r_{0}    ,\\
S(\tau) & = & \frac{2\pi}{H_0}|\sin\theta_0|(H_{0}^2|k|\tau+1)   .
\end{eqnarray}
That is, the string winds 
around the 2-sphere at the angle $\theta=\theta_0$
and expands to infinite size together with the scale factor.

Another special case is given in the limit $k = 0$ (i.e.,
$r = r_{0}$), where we have 
\begin{eqnarray}
A(\tau) & = & H_{0} |l| \tau   ,\\
B(\tau) & = & \frac{1}{H_{0}} \sqrt{1+H_{0}^4 l^2 \tau^2 }   ,\\
t(\tau) & = & \frac{1}{H_{0}} {\rm arccosh}\sqrt{1+H_{0}^4 l^2 \tau^2}   ,\\
\theta(\tau) & = & -\frac{l}{|l|}  {\rm arctg}(H^2_{0} |l| \tau) + 
\theta_{0}   ,\\
S(\tau) & = & \frac{2\pi}{H_0}| H_0^2 l \tau\sin{\theta_0}-\cos\theta_0| .
\end{eqnarray}
Notice that when $\tau$ goes from $0$ to $\infty,$ the angle $\theta$ 
changes by $\pi/2.$ Therefore we can distinguish a number of different 
scenarios: 1) $\theta_0=0.$ In this case, the string starts at the equatorial
plane and then moves towards one of the poles in such a way that its size
$S(\tau)$ is constant, i.e., the contraction of the string is exactly balanced
by the expansion of the 2-sphere. 2) $\theta_0\neq 0.$
The string starts somewhere on one hemisphere, then crosses the equator and 
approaches a fixed position on the other hemisphere. During its evolution, the
string grows indefinetely. 3) $\theta_0\neq 0.$
The string starts somewhere on one hemisphere, then moves towards the
nearest pole where it collapses. It then reappears and
approaches a fixed position on the same hemisphere. After the collapse, and
during its later evolution, its size will grow indefinetely.

Returning to the general expression (III.34), it is easy to see that the 
dynamics in the general case is qualitatively similar to the $k=0$ case,
although quantitatively different. For instance, if $\theta_0=0,$ the
string starts at a fixed angle $\theta(\tau=0)=-{\mbox{sign}}(l)
\mbox{arctg}(|k|/|l|),$ and can then approach the nearest pole in such a 
way that its size is constant. Similarly, we can also find solutions where
the string expands indefinetely, possibly after collapsing once during
its early evolution.

\subsubsection{Time-symmetric stiff-fluid solutions}

Using (III.19)-(III.20) for the null string case, we obtain from (III.27) 
\be
\tau(t) = \int_0^{t} \sqrt{\frac{M - b^2( t-\sqrt{M}/b)^2}
{l^2 b^2+k^2  M (1-b^2(t-\sqrt{M}/b)^2)/b^2}}\;dt   ,
\ee
where we took again boundary conditions such that $t(0)=0.$ From
(III.7), we get
\be 
r(\tau)=\frac{k}{b^2}\tau+r_0.
\ee
Thus $r$ is expressed directly as a function of $\tau,$ while (III.45) must
be inverted to give $t(\tau).$
Notice that (III.45) 
is an elliptic integral, which can be given in terms of elliptic functions 
\cite{as}
\bea
\tau(t)&=&\frac{b}{|k|}\sqrt{\frac{M}{b^2}+\frac{l^2b^2}{k^2}}\;E\left(\xi,
\sqrt{\frac{Mk^2}{Mk^2+l^2b^4}}\;\right)-\frac{l^2b^4}
{k^2\sqrt{Mk^2+l^2b^4}}F\left(\xi,
\sqrt{\frac{Mk^2}{Mk^2+l^2b^4}}\;\right)\nonumber\\
&-&\frac{b}{|k|}\left(\frac{\sqrt{M}}{b}-t\right)
\sqrt{\frac{M/b^2-(t-\sqrt{M}/b)^2}{M/b^2+l^2b^2/k^2-
(t-\sqrt{M}/b)^2}},
\eea
where $F,E$ are the elliptic integrals of first and second kind, 
respectively, and
\be
\xi=\mbox{arcsin}\left[\sqrt{\frac{Mk^2+l^2b^4}{Mk^2}}\;
\sqrt{\frac{M/b^2-(t-\sqrt{M}/b)^2}{M/b^2+l^2b^2/k^2-
(t-\sqrt{M}/b)^2}}\;\right].
\ee 
Here we analyse some special cases in which (III.47)  
becomes elementary. First, if 
$l = 0,$ it gives the simple solution 
\bea
t & = & \frac{|k|}{b}\tau    ,\\
r & = & \frac{k}{b^2}\tau + r_{0},\\
\theta&=&\theta_0   .
\eea
Then the string size is
\be
S(\tau)=\frac{2\pi}{b}|\sin\theta_0|\sqrt{2|k|\sqrt{M}\tau-k^2\tau^2},
\ee
that is, the string starts with zero size, then expands and eventually
recollapses together with the Universe.

Another special case is when $k = 0$, and reads as 
\bea
\tau(t) & = & \frac{M}{2|l|b^2} \left[ \left(\frac{bt}{\sqrt{M}}-1\right) 
\sqrt{1-\left(\frac{bt}{\sqrt{M}}-1\right)^2} +  
\arcsin\left(\frac{bt}{\sqrt{M}}-1\right)+\frac{\pi}{2} \right]   ,\\
r&=&r_0,\\
\theta(t) & = & \frac{l}{|l|b}\arcsin\left(\frac{bt}{\sqrt{M}}-1\right)
+\theta_0 .
\eea
In this case the string size is given by
\be
S(t)=\frac{2\pi\sqrt{M}}{b}\sqrt{1-\left(\frac{bt}{\sqrt{M}}-1\right)^2}\;
\left| \sqrt{1-\left(\frac{bt}{\sqrt{M}}-1\right)^2}\sin\theta_0\pm
\left(\frac{bt}{\sqrt{M}}-1\right)\cos\theta_0\right|.
\ee
If $\theta_0=0,$ the string is at the equator and it simply follows the 
evolution of the Universe, i.e., it starts with zero size, then expands
and eventually recollapses together with the Universe. On the other hand, if
$\theta_0\neq 0,$ the string has the possibility to pass one of the poles of 
the 2-sphere, i.e. it starts with zero size, then expands but recollapses,
then expands again and eventually recollapses together with the Universe. 

It is straightforward to check that the qualitative behaviour 
of the string solutions in the general 
case (described by III.47) essentially follows the $k=0$ case, thus we
shall not go into the quantitative details here.

\section{Strings in Bianchi I background}
\setcounter{equation}{0}

The validity of the low-energy-effective-action equations for strings in
Bianchi type homogeneous spacetimes has been studied in \cite{homoglow}, and 
in this section we consider the 
evolution of strings based on the equations (II.1)-(II.3) in Bianchi I 
background spacetimes with the metric \cite{ryan}
\be
ds^2 = dt^2 - X^2(t)dx^2 - Y^2(t)dy^2 - Z^2(t)dz^2   ,
\ee
where $X, Y, Z$ are the scale factors. 
The equations of motion and constraints are given by (II.1)-(II.3). 
Comparing with equations (III.1)-(III.2) in the case $\theta=\pi/2,$ a 
natural first attempt of an ansatz is now
\begin{equation}
X^0 = t(\tau), \hspace{0.5cm}X^1 = x = f(\tau) \cos{\sigma}, 
\hspace{0.5cm}X^2 = y =g(\tau) \sin{\sigma},
\hspace{0.5cm}X^3 = z ={\mbox{const.}}   ,
\end{equation}
and the invariant string size is 
\be
S(\tau) = \int_{0}^{2\pi} \sqrt{ f^2(\tau) X^2(t(\tau)) \sin^2\sigma + 
g^2(\tau) Y^2(t(\tau)) \cos^2\sigma}\;d\sigma   .
\ee
The ansatz (IV.2) describes a closed string of "elliptic-shape", in the sense
that
\begin{equation}
\frac{x^2}{f^2}+\frac{y^2}{g^2}=1,
\end{equation}
i.e., it generalizes the circular string ansatz considered before. This seems
to be the most natural ansatz in the spacetimes with the line element (IV.I)
because of the shear. The equations and constraints read as
\bea
\ddot{t} + XX_{,t} 
\left( \dot{f}^2 \cos^2{\sigma} - \lambda f^2 \sin^2{\sigma} \right) 
+ YY_{,t} \left( \dot{g}^2 \sin^2{\sigma} 
- \lambda g^2 \cos^2{\sigma} \right) & = & 0   ,\\
\ddot{f} + \lambda f + 2 \frac{X_{,t}}{X} \dot{t} \dot{f} & = & 0   ,\\
\ddot{g} + \lambda g + 2 \frac{Y_{,t}}{Y} \dot{t} \dot{g} & = & 0   ,
\eea
as well as
\bea
X^2 \dot{f} f - Y^2 \dot{g} g & = & 0   ,\\
\dot{t}^2 - X^2 \left( \dot{f}^2 \cos^2{\sigma} + 
\lambda f^2 \sin^2{\sigma} \right) - Y^2 \left( \dot{g}^2 \sin^2{\sigma} + 
\lambda g^2 \cos^2{\sigma} \right) & = & 0   .
\eea
Notice, however, that the equations (IV.5)-(IV.9) are not all independent.
After some algebra, 
one finds that they reduce to just four independent equations
\begin{eqnarray}
\ddot{f}+\lambda f+2\frac{X_{t}}{X}\dot{t}\dot{f}=0,\nonumber\\
\dot{t}^2=X^2\dot{f}^2+\lambda Y^2 g^2,\nonumber\\
X^2 f\dot{f}=Y^2 g\dot{g},\\
X^2\dot{f}^2+\lambda Y^2 g^2=Y^2\dot{g}^2+\lambda X^2 f^2.\nonumber
\end{eqnarray}
The last two equations of (IV.10) lead to the following two possibilities:\\

{\bf a:}
\begin{equation}
\frac{\dot{f}}{f}=\frac{\dot{g}}{g}\;\;\;\;{\mbox{and}}\;\;\;\;
X^2 f\dot{f}=Y^2 g\dot{g},
\end{equation}
which are solved by
\begin{equation}
X(t)=\pm c_1 Y(t),\;\;\;\;g(\tau)=c_1 f(\tau),
\end{equation}
where $c_1$ is a constant. After a trivial coordinate redefinition, this 
corresponds to a circular string in an axially symmetric background.\\

{\bf b:}
\begin{equation}
\frac{\dot{f}}{f}=-\lambda\frac{g}{\dot{g}}\;\;\;\;{\mbox{and}}\;\;\;\;
X^2 f\dot{f}=Y^2 g\dot{g},
\end{equation}
from which follows that
\begin{equation}
-\lambda X^2 f^2=Y^2\dot{g}^2.
\end{equation}
This equation has no real solutions for tensile strings ($\lambda=1$), 
while for null strings ($\lambda=0$), we find
\begin{equation}
f={\mbox{const.}}\equiv c_1,\;\;\;\;g={\mbox{const.}}
\equiv c_2,\;\;\;\;t={\mbox{const.}}\equiv c_3,
\end{equation}
with $X(t),\;Y(t)$ arbitrary. Such solutions, with $t$=const., have been
considered before in other contexts \cite{inigo2}, but since they do not
fulfil the physical requirement of forward propagation $(\dot{t}>0)$,
we discard them here. 

Thus our ansatz (IV.2) eventually only works in the case (IV.12). Then the 
equations (IV.10) reduce to
\begin{equation}
\ddot{f}+\lambda f+2\frac{X_t}{X}\dot{t}\dot{f}=0,
\end{equation}
\begin{equation}
\dot{t}^2=X^2\dot{f}^2+\lambda X^2 f^2.
\end{equation}
For the null strings $(\lambda=0)$, they are immediately solved by
\begin{equation}
\tau= \frac{1}{|c_1|}\int_{0}^{t} X(t) dt,
\end{equation}
\begin{equation}
f(t)=c_1\int_{t_0}^{t}\frac{dt}{X^2(t)}.
\end{equation}
For the tensile strings $(\lambda=1)$, they can not be solved in general.
However, the same equations appeared in a study of strings in 
Friedmann-Robertson-Walker Universes \cite{inigo2}, and some special 
solutions were found there.

Here we are interested in strings in Bianchi universes.
Usually, one starts with the Kasner-type vacuum power-law solutions 
\cite{landau}, which are given by 
\bea
X(t) & = & t^{p_{1}}   ,\\
Y(t) & = & t^{p_{2}}   ,\\
Z(t) & = & t^{p_{3}}   ,
\eea
and 
\bea
p_{1} + p_{2} + p_{3}  =  1   \hspace{2.cm} 
p_{1}^2 + p_{2}^2 + p_{3}^2  =  1   ,
\eea
where 
\be 
- \frac{1}{3} \leq p_{1} \leq 0  , \hspace{1.cm} 
0 \leq p_{2} \leq \frac{2}{3}  , \hspace{1.cm} 
\frac{2}{3} \leq p_{3} \leq 1   .
\ee
However, as we saw before, our string ansatz only works in axially symmetric
cases. Furthermore, we shall usually also allow the presence of matter.

A special case of the model (IV.1) is an axially symmetric Kasner model in
which the matter is that of the stiff-fluid. The metric reads   
\be
ds^2 = dt^2 - A^2(t)\left( dx^2 + dy^2 \right) - Z^2(t)dz^2   ,
\ee
and it is just Kantowski-Sachs metric (III.1) of zero curvature. 
For the stiff-fluid, $p = \rho,$ the conservation law is given by 
$\rho Z^2 A^4 = \rho t^2 = k^2/16\pi,$ which gives the 
solutions for the scale factors in the form \cite{axkasner} 
\bea 
A(t) & = & t^{p_{A}}   ,\\
Z(t) & = & t^{1 - 2p_{A}}   ,
\eea
where
\bea
p_{A}=\frac{1}{3}[1\pm\sqrt{1-3k^2/2}\;],
\eea
that is, $0\leq p_{A}\leq 2/3$.

Under the ansatz (IV.2) with $f=g,$ the equations of motion 
(IV.16)-(IV.17) become 
\bea
\ddot{f} +  \lambda f+\frac{2p_{A}}{t}\dot{t}\dot{f} & = & 0   ,\\
\dot{t}^2-t^{2p_{A}}(\dot{f}^2+\lambda f^2) & = & 0   .
\eea

\subsection{Tensile strings.}

For the tensile strings ($\lambda = 1$), equations (IV.29)-(IV.30) 
were considered in
\cite{inigo2}. They do not seem to be integrable, 
but some special solutions were
found
\be
t(\tau)  =  A\exp{(c_1\tau)}   ,\;\;\;\;
f(\tau)  =  B \exp{(c_2\tau)}   ,
\ee
where the constants $(A,B,c_1,c_2)$ are given by
\be
c_1=\frac{\mp 1}{\sqrt{(p_{A}-1)(p_{A}+1)}},\;\;\;\;
c_2=\pm\sqrt{\frac{p_{A}-1}{p_A+1}},\;\;\;\;
B=\frac{A^{1-p_{A}}}{\sqrt{2p_{A}(p_{A}-1)}}.
\ee
However, this solution is not real for the values allowed in our case 
($0 \leq p_{A} \leq 2/3$),
so it must be discarded.
 
All we can do then is to  determine the  asymptotics of 
the solutions to equations (IV.29)-(IV.30). One finds for $\tau \to
\infty$
\bea
t(\tau) & = & A\tau   ,\\
f(\tau) & = & A^{1-p_{A}} \tau^{-p_{A}} \cos{\tau}   ,
\eea
where $A$ is an arbitrary positive constant.
The invariant string size reads as ($\tau \to \infty$) 
\be
S(\tau) = 2\pi A\vert \cos{\tau} \vert  ,
\ee
so it asymptotically oscillates with 
constant amplitude and unit frequency, while the comoving string size goes to
zero. 

\subsection{Null strings.}

For the null strings ($\lambda = 0$) in axially symmetric Kasner spacetime, 
the equations (IV.18)-(IV.19) are integrated to give
\bea
t(\tau) & = & \left[\vert c_1 \vert (1+p_{A}) 
\tau\right]^{\frac{1}{1+p_{A}}}   ,\\
f(\tau) & = & c_1 \left( \frac{1 + p_{A}}{1 - p_{A}} \right)
\left[\vert c_1 \vert (1+p_{A}) \tau\right]^{\frac{1 - p_{A}}{1 + p_{A}}}   .
\eea
In this case the invariant string size is 
\be
S(\tau) = 2\pi \vert c_1 \vert \left(\frac{1+p_{A}}{1-p_{A}} \right)
\left[\vert c_1 \vert (1+p_{A})\tau\right]^{\frac{1}{1 + p_{A}}}   ,
\ee
which blows up for $\tau \to \infty.$ This is also the case for the
comoving string size.
\vskip 16pt
We close this section with some comments on the possibility of having
elliptic-shaped strings in anisotropic Bianchi backgrounds. As we saw,
the ansatz (IV.2) led to inconsistencies unless $f(\tau)=g(\tau)$ and
$X(t)=Y(t).$ However, this does not mean that we must completely rule
out the possibility of having elliptic-shaped string configurations.
In fact, it is possible to make an ansatz more general than (IV.2),
but still describing an elliptic-shaped string. This can be done along
the lines of the procedure used in reference \cite{china} (in a 
somewhat different context): We discard the orthonormal gauge and work
directly with the Nambu-Goto action. In that case, the ansatz (IV.2)
leaves us more freedom than before. Unfortunately, the equations of motion now
become more complicated than before, but at least they are not
explicitly inconsistent, and there is some hope that one can find
special solutions or at least solve the equations numerically.

In the orthonormal gauge (II.2)-(II.3), as used in this paper, this more 
general ansatz corresponds to replacing (IV.2) by
\be
X^0=t(\tau),\hspace{0.5cm}X^1=x=f(\tau)\cos\phi(\tau,\sigma),
\hspace{0.5cm}X^2=y=g(\tau)\sin\phi(\tau,\sigma),\hspace{0.5cm}
X^3=z={\mbox{const}}.,
\ee
and the function $\phi(\tau,\sigma)$ gives us the extra freedom as mentioned
above. However, we leave the implications of using the ansatz (IV.39)
for investigations elsewhere.

\section{Strings in Axisymmetric Bianchi type IX Background}
\setcounter{equation}{0}

Another interesting example of a curved background for strings, we consider, 
is the Bianchi type IX background. It generalizes 
the $k = +1$ isotropic Friedmann 
model to the case of anisotropic spacetimes. In order to show the relation, it
has been shown, among others, that all BIX models recollapse similarly as 
$k = +1$ Friedmann models \cite{wald}. The general case cannot be solved 
analytically for the scale factors and they subject to chaotic behaviour. It
would be interesting to find out whether the test strings in such a general
background also behave chaotically, but for the moment we leave this question
for a separate paper, and consider just an axially symmetric Bianchi IX model
which can be solved analytically. The metric of such a model in a holonomic 
frame, is given by \cite{brill,j2} 
\be
ds^2 = dt^2 - c^2(t) \left( d\psi + \cos{\theta}d\varphi \right)^2 - 
a^2(t) \left( d\theta^2 + \sin^2{\theta} d\varphi^2 \right)   ,
\ee
where $\psi, \theta, \varphi$ are the Euler angles ($0 \leq \psi 
\leq 4\pi, 0 \leq \theta \leq \pi$ and $0 \leq \varphi \leq 2\pi$). 
We use the following ansatz for the spacetime coordinates
\be
X^0 = t(\tau), \hspace{0.5cm}X^1 
= \psi(\tau), \hspace{0.5cm}X^2 = \theta(\tau),
 \hspace{0.5cm}X^3 = \varphi = \sigma,   
\ee
and the equations of motion (II.1) then read  
\bea
\ddot{t} + cc_{,t}\dot{\psi}^2 + aa_{,t} \dot{\theta}^2 - 
\lambda \left( cc_{,t}\cos^2{\theta} + 
aa_{,t}\sin^2{\theta} \right)& = & 0   ,\\
\ddot{\psi} + 2 \frac{c_{,t}}{c} \dot{t}\dot{\psi} + \frac{c^2}{a^2}
\cot{\theta} \dot{\psi} \dot{\theta} & = & 0   ,\\
\ddot{\theta} + 2 \frac{a_{,t}}{a} \dot{t}\dot{\theta} - 
\lambda \frac{c^2 - a^2}{a^2} \sin{\theta}\cos{\theta} & = & 0   ,\\
\frac{c^2}{a^2} \frac{1}{\sin{\theta}} \dot{\psi} \dot{\theta} & = & 0   .
\eea
The last of these equations 
(V.6) says that either $\psi =$ const. or $\theta =$
const. ($\theta \neq 0$). The constraints (II.2)-(II.3) read 
\bea
\dot{t}^2 - c^2 \dot{\psi}^2 - a^2 \dot{\theta}^2 - \lambda \left( 
a^2 \sin^2{\theta} + c^2 \cos^2{\theta} \right) & = & 0   ,\\
c^2 \cos{\theta} \dot{\psi} & = & 0   ,
\eea
from which it follows that either 
$\psi$ must be constant or $\theta = \pi/2$. We will
consider both cases. For each of these cases the invariant string size is given
by 
\be
S(\tau) = 2\pi \sqrt{a^2(t(\tau)) \sin^2{\theta(\tau)} + c^2(t(\tau))
\cos^2{\theta(\tau)}}   .
\ee
The well-known stiff-fluid solutions of the Einstein equations for Bianchi IX
axisymmetric model are given by 
\cite{j2}
\bea
c^2 & = & \frac{A}{\cosh{A\eta}}   ,\\
a^2 & = & \frac{B^2 \cosh{A\eta}}{4A \cosh^2{( \frac{B}{2}
\eta)}}   ,\\
p & = & \rho = \frac{M^2}{4a^4c^2}   ,
\eea
where $A, B, M$ are constants $( A > \vert B \vert )$ with 
\bea
B^2 & = & A^2 - M^2      ,\\
t & = & \int^{\eta} a^2\vert c \vert d\eta   .
\eea
One can easily see that the vacuum solution $M = 0$ is given by 
\bea
c^2 & = & \frac{A}{\cosh{A\eta}}   ,\\
a^2 & = & \frac{A \cosh{A\eta}}{4 \cosh^2{ (\frac{A}{2}\eta)}}   .
\eea
Notice that the scale factor $c(\eta)$ increases from $c(-\infty)=0$ to
$c(0)=c_{\mbox{max}}=\sqrt{A}$ and then decreases to $c(\infty)=0$
again. On the other hand, the scalefactor $a(\eta)$ decreases from
$a(-\infty)=\sqrt{A/2}$ to $a(0)=a_{\mbox{min}}=\sqrt{A/4}$ and then
increases to $a(\infty)=\sqrt{A/2}$ again. However, the volume 
essentially follows the scalefactor $c(\eta)$, i.e., the 
universe is of recollapsing type.

\subsection{$\theta = \pi/2$ solutions}

If we assume $\theta = \pi/2$ and $\dot{\psi} \neq 0$, then we can easily
integrate (V.4) to give 
\be
\dot{\psi} = \frac{m}{c^2}  ,
\ee
with $m =$ const. Then we have 
\bea
\tau(t) & = & \int_{0}^{t}
{\frac{\vert c \vert dt}{\sqrt{m^2 + \lambda a^2 c^2}}}   ,\\
\psi(t) & = & m \int_{t_0}^{t}
{\frac{dt}{\vert c \vert \sqrt{m^2 + \lambda a^2 c^2}}}
  .
\eea
For the null strings $(\lambda=0),$ 
we can easily get the solution for the vaccum case 
$M = 0$ in terms of parametric time $\eta,$ by using (V.15) and (V.16), i.e., 
\be
d\eta = \frac{\vert m \vert}{a^2 c^2}d\tau   ,
\ee
which is integrated to
\be
\eta(\tau) = \frac{2}{A} {\rm arcth} \left( \frac{2\vert m \vert}{A} 
\tau \right)   ,
\ee
where we took boundary conditions such that $\eta(0)=0.$
The solutions for the scale factors $c(\tau)$ and $a(\tau)$ are given by 
\bea
c^2(\tau) & = &  A \left(\frac{A^2 - 4m^2 \tau^2}{A^2 +
4m^2 \tau^2}\right)   ,\\
a^2(\tau) & = & \frac{1}{4A} \left( A^2 + 4m^2 \tau^2 \right)   ,
\eea
and $\tau^2 \leq A^2/4m^2$.
Then we find 
\bea
t(\tau) & = & \frac{\vert m \vert}{\sqrt{A}}
\int_{0}^{\tau} d\tau \sqrt{\frac{A^2+4m^2\tau^2}{A^2-4m^2\tau^2}} ,\\
\psi(\tau) & = & \psi_{0}+\frac{m}{A}  
\int_{0}^{\tau} d\tau \left(\frac{A^2+4m^2\tau^2}{A^2-4m^2\tau^2}\right).
\eea
The integral for $t(\tau)$ is of elliptic type, while the one for $\psi(\tau)$
is elementary. However, we shall not need the explicit results here.

The invariant string size reads
\be
S(\tau) = \frac{\pi}{\sqrt{ A }} \sqrt{A^2 + 4m^2\tau^2}  .
\ee
From the above, we conclude that for the admissible values of the parameter
$\tau,$ the string starts with the size $S = \pi \sqrt{2 A}$ for 
$\tau  = - A/2\vert m\vert,$ then contracts to the size 
$S = \pi \sqrt{ A }$ for $\tau = 0,$ and expands again to 
$S = \pi \sqrt{2 A }$ for $\tau =  A/2\vert m \vert$. This can be easily
understood physically since, for $\theta=\pi/2,$ the string is winding 
in the $\varphi$-direction with scale factor $a.$

\subsection{The case $\dot{\psi} = 0, \dot{\theta} \neq 0$}

In this case, the equations (V.3)-(V.8) become 
\bea
\ddot{t} + aa_{,t} \dot{\theta}^2 - 
\lambda \left( cc_{,t}\cos^2{\theta} + 
aa_{,t}\sin^2{\theta} \right)& = & 0   ,\\
\ddot{\theta} + 2 \frac{a_{,t}}{a} \dot{t}\dot{\theta} - 
\lambda \frac{c^2 - a^2}{a^2} \sin{\theta}\cos{\theta} & = & 0   ,\\
\dot{t}^2 - a^2 \dot{\theta}^2 - \lambda \left( 
a^2 \sin^2{\theta} + c^2 \cos^2{\theta} \right) & = & 0   .
\eea
Notice that the first equation can be obtained from the two others. Thus
we have just two coupled equations; one of first order and one of
second order. For the tensile strings, the general solution does not 
seem to be available.
For the null strings, we can integrate for $\theta(\tau)$ and $t(\tau)$ 
\bea
\dot{\theta} & = & \frac{s}{a^2}   ,\\
\dot{t}^2 & = & \frac{s^2}{a^2}   ,
\eea
with $s =$ const.
Then using the exact vacuum solutions (V.15)-(V.16), we can integrate this
further since $(d\eta/d\tau)^2 = s^2/a^6c^2$. One finds
\be
\vert s \vert d\tau = \frac{A^2}{8}
\frac{\cosh{A\eta}}{\cosh^3\left(\frac{A}{2} \eta \right)} d\eta   ,
\ee
which can be integrated explicitly in terms of elementary functions.
In principle we can then also obtain expressions for $t(\tau)$ and
$\theta(\tau).$ However, it turns out to be somewhat simpler to express 
everything in terms of the parametric time $\eta.$ For instance
\be
d\theta=\frac{s}{\vert s \vert} \vert ac \vert d\eta,
\ee
which leads to
\be
\theta(\eta)-\theta_0=2\frac{s}{\vert s \vert}
{\mbox{arctg}}\left( e^{A\eta/2}\right).
\ee
It is then straightforward to write down an explicit expression for
the invariant string size similar to equation (V.9), 
but with $S$ as a function of $\eta,$ since the scale factors
are already given in terms of $\eta.$

It follows from the above results 
that during the whole evolution of the Universe, the polar angle
$\theta$ changes by $\pi.$ Thus there are two scenarios: If $\theta_0=0,$ then
the string starts with zero size at one of the poles, then expands and 
eventually collapses to zero size again at the other pole. On the other hand,
if $\theta_0\neq 0,$ then the string starts with finite size, passes one of the
poles (still with finite size) and eventually ends up with a finite size.
Thus the behaviour is qualitatively similar to that of strings in 
Kantowski-Sachs spacetimes as described in Section III.  

\section{Summary}

In this paper we have considered the tensile and null string evolution
and propagation in some homogeneous but anisotropic spacetimes of 
Kantowski-Sachs and Bianchi type. This generalizes and completes
earlier investigations of strings in more symmetric backgrounds.

Our results demonstrate the richness of different evolution schemes for
extended objects, here strings, in curved backgrounds. For the tensile strings,
this is due to the "competition" between the string tension and the
gravitational field, which together determine the evolution of the string.
For the null strings, it is simply due to the fact that we are dealing
with an extended object in a gravitational field, i.e., the object  
subjects to tidal forces. In both cases, the situation should be compared with
the conceptually much simpler problems of point particle propagation in curved 
spacetimes and string evolution and propagation in flat Minkowski spacetime.

We mainly considered closed circular strings, which allowed us to obtain simple
exact analytical results in most cases. We essentially saw three qualitatively
different kinds of circular string evolution: a) the string simply follows
the expansion or contraction of the Universe, b) the string makes a finite or
infinite number of oscillations during the evolution of the Universe, c) the
contraction of the string is exactly balanced by the expansion of the Universe,
such that the physical string size is constant.

We also discussed the problems of obtaining consistent equations of motion
describing an elliptic-shaped string configuration. The use of the simplest, 
and a priori most natural ansatz, describing an elliptic-shaped string, led to
inconsistent equations of motion. At this point, we leave for future work
the question of whether a more complicated ansatz, as described at the end of
Section IV, with its more complicated equations of motion,  can solve these 
problems.

\section{Acknowledgments}
We wish to thank A.A.Zheltukhin for useful discussions. 
MPD is supported by the Polish Research Committee (KBN) grant No 2 PO3B 196 10.
The work of ALL  was partly supported by NSERC (Canada) and CNRS (France).

\end{document}